\documentclass[structabstract]{aa}
\usepackage{graphicx}
\usepackage{txfonts}
\begin{document}
   \title{The progenitor of binary millisecond radio pulsar
 PSR J1713+0747 (Research Note)}

   \author{Wen-Cong Chen
     \inst{1,}
     \inst{2,}
     \inst{3}
     \and
     Jorge A. Panei
     \inst{4,}
     \inst{5}
      }

   \institute{Department of Physics, Shangqiu Normal University,
     Shangqiu 476000, China
     \and
     School of Physics and State Key Laboratory of Nuclear Physics
     and Technology, Peking University, Beijing 100871, China
     \and
     Key Laboratory of Modern Astronomy and Astrophysics (Nanjing University), Ministry of Education, Nanjing 210093, China
     \\
     \email{chenwc@nju.edu.cn}
     \and
     Facultad de Ciencias Astron\'omicas y Geof\'{\i}sicas, UNLP,
     Paseo del Bosque S/N, La Plata B1900FWA, Argentina
     \and
     Instituto de Astrof\'{\i}sica de La Plata (CCT La Plata),
     CONICET-UNLP, Argentina\\
     \email{panei@fcaglp.unlp.edu.ar}
     }

   \date{Received ...; accepted ...}


   \abstract
   {PSR J1713+0747 is a binary system comprising millisecond radio pulsar with a spin period of
     4.57 ms, and a low-mass white dwarf (WD) companion orbiting the pulsar with a period of 67.8 days. Using the general relativistic Shapiro
     delay, the masses of the WD and pulsar components were previously found to be $0.28\pm 0.03 M_{\odot}$ and $1.3\pm
     0.2 M_{\odot}$ (68\% confidence), respectively.}
   {Standard binary evolution theory suggests that PSR J1713+0747 evolved from a low-mass X-ray binary (LMXB). Here, we test this hypothesis.}
   {We used a binary evolution code and a WD evolution code to
     calculate evolutionary sequences of LMXBs that could result
     in binary millisecond radio pulsars such as PSR
     J1713+0747. }
   {During the mass exchange, the mass transfer is
     nonconservative. Because of the thermal and viscous
     instabilities developing in the accretion disk, the neutron star accretes only a small part of
     the incoming material. We find that the progenitor of PSR
     J1713+0747 can be modelled as an LMXB including a donor star
     with mass $1.3-1.6 M_{\odot}$ and an initial orbital period
     ranging from 2.40 to 4.15 days. If the cooling timescale of the WD is 8
     Gyr, its present effective temperature is between 3870 and 4120 K, slightly higher than
     the observed value. We estimate a surface gravity of ${\rm
       Log} (g) \approx 7.38 - 7.40$.}
   {}

\keywords{binaries: general -- stars: evolution -- stars: white
dwarfs -- pulsars: individual (PSR J1713+0747)}

     \titlerunning{The progenitor of PSR J1713+0747}
\authorrunning{Chen \& Panei }

   \maketitle


\section{Introduction}

Since the first X-ray source Sco X-1 was discovered
(\cite{giac62}), it was proposed that the strong Galactic X-ray
sources were accreting neutron stars or black holes in binary
systems (\cite{zel66,novi66,shkl67}). These sources have
significant higher flux in X-ray than other wavelengths. Over 90\%
of strong Galactic X-ray sources can be classified in two kinds as
follows: high-mass X-ray binaries and low-mass X-ray binaries
(LMXBs). The latter consists of a neutron star or a black hole and
a low-mass donor star ($\la 1-2 M_{\odot}$).

The mechanisms that drive mass transfer in LMXBs depend on the
initial separations of the binary components (\cite{bhat91}). In
narrow systems with an orbital period ($P_{\rm orb}$) less than
two days, mass transfer is driven by angular momentum loss due to
gravitational radiation and magnetic braking. However, mass
transfer is driven by the evolution of the secondary in relatively
wide LMXBs ($P_{\rm
  orb}\ga 2$ d). In LMXBs, mass accretion onto neutron star causes a
magnetic field decay and spins it up to a short spin period ($\la
20$ ms).  It is believed that circular binaries including
millisecond  radio pulsar and He or CO white dwarfs (WDs) are
formed as a result of long-term mass accretion
(\cite{alpa82,stai04}). The discovery of the first millisecond
X-ray pulsar in LMXB SAX 1808.4-3658 strongly supports this
scenario (\cite{wijn98}).

Through a survey for millisecond pulsars with the 305 m Arecibo
radio telescope, \cite{fost93} have discovered PSR J1713+0747.
This pulsar has a spin period of 4.57 millisecond. It is in a
circular orbit of 67.8 days and has a low-mass WD as companion. It
has a weak magnetic field of $1.9\times 10^{8}$ G and a spin-down
age of $8.9\pm 1.0$ Gyr, so is an older neutron star. By detecting
the general relativistic Shapiro delay, the neutron star and WD
were set up as the low limits for the masses of $1.2 M_{\odot}$
and $0.27 M_{\odot}$, respectively (\cite{cami94}).

Through observations for 12 years between April 1992 and May 2004,
343 pulse times of arrival for PSR J1713+0747 were recorded with
uncertainties down 200 nanosecond. These timing data can result in
the measurement of the Shapiro delay, which can constrain the mass
of the WD $M_{2}$ and the inclination of the orbit $i$. The
observational data constrained the masses of the pulsar to $1.3\pm
0.2 M_{\odot}$ and the WD
 to $0.28\pm 0.03 M_{\odot}$ (68\%
confidence, \cite{spla05}).

It is clear that PSR J1713+0747 is a recycled pulsar, because it
gained mass and angular momentum from its donor star, the
progenitor of the WD (\cite{alpa82}). Assuming the mass of the WD
is equal to the helium core mass of its progenitor, binary
evolution theory has presented a peculiar relation between the
orbital period $P_{\rm orb}$ and the mass of WD $M_{2}$
(\cite{rapp95,taur99,pods02}). According to this relation, Tauris
\& Savonije (1999) obtained the range of the WD mass for PSR
J1713+0747, $0.31 M_{\odot}<M_{2}<0.34 M_{\odot}$. Later on,
$M_{2}$ was constrained in the range of $0.30 - 0.35 M_{\odot}$
(\cite{pfah02}). After adopting the predicted value by the
theoretical relation of $P_{\rm orb}-M_{2}$, the pulsar has a
somewhat higher mass, $M_{1}=1.53^{+0.08}_{-0.06}M_{\odot}$ (68\%
confidence) (\cite{spla05}).

Assuming that the neutron star accretes 10 \% of the material
coming from the donor star, \cite{benv06} have calculated the
evolution of PSR J1713+0747 progenitor to fit its component masses
and the orbital period. Their results show that the progenitor of
PSR J1713+0747 may be an LMXB with initial masses of
$M_{1}=1.4M_{\odot}$ and $M_{2}=1.5M_{\odot}$ and an initial
orbital period of $P_{\rm i}=3.10$ d (Z=0.02) or $P_{\rm i}=3.05$
d (Z=0.01). Adopting a semi-analytic approach, Chen et al. (2006)
have also performed binary evolution calculations to account for
the component masses and the orbital period of PSR J1713+0747.

In this paper, where we employ a detailed stellar evolutionary
code and a WD evolutionary code, we attempt to simulate the
evolutionary history of the progenitor of binary millisecond radio
pulsar PSR J1713+0747. The structure of this paper is organized as
follows. In section 2, we describe the stellar evolutionary codes,
input physics in the binary evolution calculations, and the WD
evolutionary code. Numerically calculated results for the
evolutionary sequences of binary system resulting in the birth of
PSR J1713+0747 are presented in section 3. Finally, we discuss and
summarize our results in section 4.

\section[]{Stellar evolutionary code and input physics}

By employing an updated version of the stellar evolutionary code
developed by \cite{eggl71,eggl72} (see \cite{han94,pols95}), we
investigated the evolution of LMXBs. The LMXBs consist of a main
sequence donor star with a mass of $M_{2}$ and a neutron star with
a mass of $M_{1}$. Radiative opacities are those from
\cite{roge92} and at low-temperature molecular opacities from
\cite{alex94}. For the secondary star, we take a mixing-length
parameter of $\alpha=2.0$ and an effective Roche lobe radius given
by \cite{eggl83}
\begin{equation}
\frac{R_{\rm L}}{a}=\frac{0.49q^{2/3}}{0.6q^{2/3}+ {\rm
ln}(1+q^{1/3})},
\end{equation}
where $a$ and $q = M_{2}/M_{1}$ are the orbital separation and the
mass ratio of the binary, respectively.

The mass transfer rate from the donor star via Roche lobe overflow
can be written as
\begin{equation}
\dot{M}_{2}=-{\rm RMT}\cdot {\rm max}[0,(R_{2}/R_{\rm L}-1)^{3}]
\emph{M}_{\odot}\rm yr^{-1},
\end{equation}
here we take RMT to be 1000 in the calculations. We neglect the
spin angular momentum of LMXBs because it is smaller compared with
the total orbital angular momentum of the system \footnote{If we
take typical parameters for the binary, such as
$M_1=1.4~M_\odot,M_2=1.0~M_\odot,R_1=10^6~{\rm cm},
R_2=1~R_\odot$, the spin period of the NS is 1 ms, and the orbital
period is $P_{\rm orb}=$ 3 d, the orbital angular momentum of the
binary is estimated to be $J_{\rm orb}=1.9\times10^{52}~{\rm
g\,cm^2s^{-1}}$. Assuming that the donor star corotates with the
orbital rotation, we have $J_1=7.0\times10^{48}~{\rm
g\,cm^2s^{-1}}$, $J_2=2\pi k^{2}M_{2}R_{2}^{2}/P_{\rm
orb}=2.4\times10^{49}~{\rm g\,cm^2s^{-1}}$ ($k^{2}\approx 0.1$ for
main-sequence stars, \cite{derv10}), where $J_1, J_2$ are the spin
angular momentum of the NS, and the donor star, respectively. It
is clear that the orbital angular momentum is three to four orders
of magnitude greater than the spin angular momentum. }. In the
calculations we consider two types of mechanisms for orbital
angular momentum loss from the binary system, they are described
as follows.

\subsection{Orbital angular momentum loss mechanisms}

\subsubsection{Magnetic braking}

The low-mass main-sequence donor stars in LMXBs were braked by the
coupling between the magnetic field and the stellar winds
(\cite{verb81}). The loss of specific angular momentum in stellar
winds is very large because the outflowing ionized particles are
bound in the magnetic field lines to corotate with the stars out
to the magnetospheric radius (see \cite{webe67,mest87,kalo99}). In
LMXBs, the spin angular momentum loss of the donor star by
magnetic braking causes spin down, while the tidal interaction
between two components would continuously act to spin the donor
star back up into corotation with the orbital motion. Because the
spin-up of the donor star consumes the orbital angular momentum,
magnetic braking indirectly extracts the orbital angular momentum
of LMXBs.

A standard magnetic braking model for the angular momentum loss is
widely applied in studying the evolution of cataclysmic variables,
and the angular momentum loss rate can be written as
(\cite{rapp83})
\begin{equation}
\dot{J}_{\rm MB}\simeq -3.8\times
10^{-30}M_2R_{\odot}^4(R_2/R_{\odot})^{\gamma}\omega^3\,{\rm
dyn\,cm},
\end{equation}
where $R_{2}$ is the radius,  $\omega$ the angular velocity of the
donor star, and $\gamma$ is a dimensionless parameter in the range
of zero to four.

However, observations based on rapidly rotating stars with a spin
period less than two to five days in young open clusters, indicate
that the standard magnetic braking description should overestimate
the angular momentum loss rate (\cite{quel98,andr03}).
\cite{sill00} propose the induced magnetic braking model, and the
angular momentum loss rate is given by
\begin{equation}
\dot{J}_{\rm mb}= \left\{\begin{array}{l@{\quad}l} -K\omega
^{3}\left(\frac{R_2}{R_{\odot}}\frac{M_{\odot}}{M_2}\right)^{1/2},
& \omega\leq\omega_{\rm crit} \strut\\
-K\omega \omega_{\rm
crit}^{2}\left(\frac{R_2}{R_{\odot}}\frac{M_{\odot}}{M_2}\right)^{1/2}
, & \omega>\omega_{\rm crit} \strut\\\end{array}\right.
\end{equation}
where $K=2.7\times 10^{47} \rm g\, cm^{2}$ (\cite{andr03}).
Observations show that there is a saturation level in the angular
momentum loss, and $\omega_{\rm crit}$ is the critical angular
velocity at which the angular momentum loss rate reaches a
saturated state. \cite{kim96} suggest that $\omega_{\rm crit}$ is
inversely proportional to the convective turbulent timescale in
the star at 200 Myr age,
\begin{equation}
\omega_{\rm crit}=\omega_{\rm
crit,\odot}\frac{\tau_{\odot}}{\tau},
\end{equation}
where $\omega_{\rm crit,\odot}=2.9\times 10^{-5}$ Hz,
$\tau_{\odot}$, and $\tau$ are the convective turbulent timescales
of the Sun and donor star, respectively (\cite{kris97}).

\subsubsection{Accretion disk instability and mass loss}

According to the mass of PSR J1713+0747 inferred by its Shapiro
delay, there is a large amount of transfer material that should be
lost from this binary system during the mass exchange. Either the
accretion disk instabilities (\cite{prin81,para96}) or the
propeller effect (\cite{illa75}) may be responsible for the mass
loss even for sub-Eddington accretion. During the mass transfer
via Roche Lobe overflow, the transferring material would form an
accretion disk around the neutron star. It is well known that disk
accretion in LMXBs may suffer thermal and viscous instability, if
the effective temperature in the disk is below the hydrogen
ionization temperature about 6500 K (\cite{para96,king97,laso01}).
This corresponds to a critical mass-transfer rate under which the
accretion disk is unstable.  This critical mass-transfer rate is
(\cite{para96,dubu99})
\begin{equation}
\dot{M}_{\rm cr} \simeq
3.2\times10^{-9}\left(\frac{M_{1}}{1.4M_{\odot}}\right)^{0.5}
\left(\frac{M_{2}}{1M_{\odot}}\right)^{-0.2}
\left(\frac{P_{\rm orb}}{1 \rm d}\right)^{1.4}M_{\odot}{\rm yr}^{-1}.
\end{equation}

When an accretion-disk instability occurs, the accreting neutron
star will become a transient X-ray source, experiencing
short-lived outbursts separated by long-term quiescence. When the
mass transfer rate $-\dot{M}_{2}$ is lower than $\dot{M}_{\rm
cr}$, we assume that the neutron star accretes only during
outbursts, so the accretion rate is $\dot{M}_{\rm ac}t_{\rm
out}=-\dot{M}_{2}t_{\rm r}$, where $t_{\rm out}, t_{\rm r}$ are
the outburst timescale and the recurrence time, respectively.
 Introducing the duty cycle $f=t_{\rm out}/ t_{\rm r}$, we have
$\dot{M}_{\rm ac}=-\dot{M}_{2}/f$ [a typical duty cycle is in the
range of 0.1-0.001
  (\cite{king03})]. If  $-\dot{M}_{2}\ge\dot{M}_{\rm
cr}$, the accretion rate is $\dot{M}_{\rm
  ac}=-\dot{M}_{2}$. In addition, in both cases the mass growth rate
of the neutron star $\dot{M}_{1}$ is limited to the Eddington
accretion rate $\dot{M}_{\rm Edd}$, i. e.
\begin{equation}
\dot{M}_{1}=\rm min [\dot{M}_{\rm ac},\dot{M}_{\rm Edd} ],
\end{equation}
where $\dot{M}_{\rm Edd}\approx1.5\times 10^{-8}M_{\odot}{\rm
yr}^{-1}$ is the Eddington accretion rate of the neutron star.

The transferred matter that exceeds the Eddington accretion rate
is assumed to be ejected in the vicinity of the neutron star. The
matter forms an isotropic wind and carries away the specific
orbital angular momentum of the neutron star. Based on the above
assumptions, we can derive the angular momentum loss rate:

\begin{equation}
\dot{J}_{\rm IW}=\frac{\dot{M}M_{2}}{MM_{1}}J,
\end{equation}
where $ J=a^{2}\mu\Omega $ is the total orbital angular momentum
of the LMXB with a circular orbit,
$\mu=M_{1}M_{2}/{(M_{1}+M_{2})}$, $M=M_{1}+M_{2}$, and
$\dot{M}=\dot{M}_{1}-\dot{M}_{\rm ac}$ are the reduced mass, the
total mass, and the mass-loss rate of the binary system,
respectively.

\subsection[]{WD regime: the evolutionary code}

When the LMXB becomes a detached system, the donor star evolves
into an He-core WD. From this stage, during thermonuclear flashes
scenario and in WD regime, we adopt the stellar evolutive code
{\tt LPCODE} (\cite{alth05}). The code is based on a detailed
description of the main physical processes involved in the
formation of WD. Radiative opacities are those from OPAL
(including carbon- and oxygen-rich compositions) for arbitrary
metallicity (Iglesias \& Rogers 1996). These opacities are
calculated for metallicities consistent with the diffusion
predictions. During the WD cooling regime gravitational settling
leads to metal-depleted outer layers. The equation of state is an
updated version of that of \cite{magn79}.  High-density conductive
opacities and the various mechanisms of neutrino emission are
taken from works of \cite{itoh92}. Hydrogen burning is taken into
account by considering by a complete network of thermonuclear
reaction rates corresponding to the proton-proton chain and CNO
cycle. Nuclear reactions rates are taken from \cite{caug88}. We
consider the evolution of the chemical abundance distribution
caused by element diffusion during the WD regime. Our treatment of
time-dependent diffusion is based on the multicomponent gas
treatment presented by \cite{burg69}, for element diffusion scheme
we considered the following nuclear species $^1$H, $^3$He, $^4$He,
$^{12}$C, $^{14}$N, and $^{16}$O. Abundance changes are computed
according to element diffusion and then to nuclear reactions and
convective mixing.

In this work, we have computed the complete evolution of seven
remnants arisen from binary systems as soon as the transfer of mass
has ceased. The masses of these seven pre-WDs are listed in Table 1
($M_{\rm 2,f}$). The evolution of these WDs is continued until they
arrive at a luminosity of ${\rm Log}(L/L_{ \odot}) \approx -4.8$.

\section[]{Results}

To study the characteristics of the progenitor of PSR J1713+0747,
we calculated evolutionary sequences of many LMXBs. We consider
both the angular momentum loss mechanisms described in section
2.2, adopting the induced magnetic braking model (eq [4]) for the
donor stars with mass less than $1.5M_{\odot}$. Some initial input
parameters were set as follows, the neutron star has a canonical
mass of $M_{1,\rm i}=1.4M_{\odot}$, the mass of the donor star
$M_{2,\rm
  i}$ is in the range of $1.0-2.0M_{\odot}$ (with a solar metallicity
of Z=0.02), the duty cycle $f=0.01$.

The characteristic age of PSR J1713+0747 estimated by Splaver et
al. (2005) is $\tau_{\rm PSR}\approx 8$ Gyr. This implies that the
WD has a cooling timescale $\tau_{\rm cool}$ of approximately 8
Gyr since the end of the mass transfer episode. Based on different
initial orbital period $P_{\rm orb,i}$, we stop the binary
evolution calculation when the binary becomes a detached system
and the donor star evolve into a WD. Subsequently, a WD
evolutionary code ({\tt LPCODE}) is applied to the donor star.  If
the final donor star mass is in the range of $0.25$ to $0.31
M_{\odot}$, and the orbital period of the binary is approximately
of $67.8$ d, the corresponding binary was assumed to be the
progenitor candidate of PSR J1713+0747.

\begin{figure}
\centering
\includegraphics[angle=0,width=9cm]{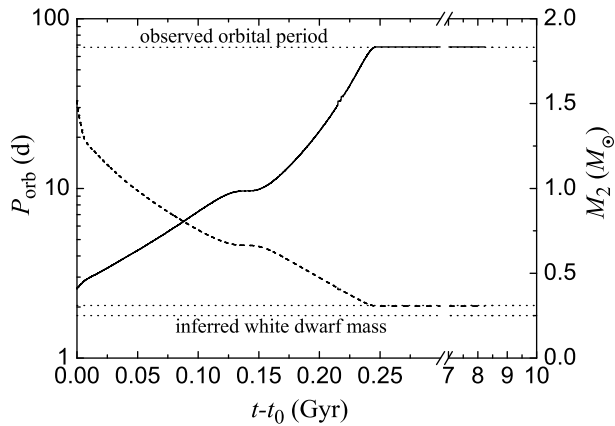}
\caption{Evolution of the orbital period (solid curve) and the donor
  star mass (dashed curve) vs. evolutive timescale for LMXB. LMXB has
  a donor star with an initial mass of $1.5M_{\odot}$ and an initial
  orbital period of 2.57 d. Horizontal dotted lines on the bottom of
  the figure correspond to the observed mass of WD with its
  error, and horizontal dotted line on the top of the figure
  represents the observed orbital period. }
\label{Fig1}
\end{figure}

\begin{figure}
\centering
\includegraphics[angle=0,width=9cm]{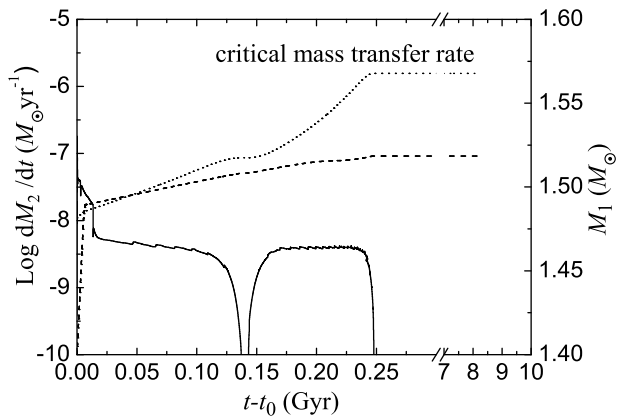}
\caption{Evolution of the mass transfer rate (solid curve) and the
  neutron star mass (dashed curve) with the evolutive timescale for
  the same LMXB of Fig. \ref{Fig1}. The dotted curve represents the
  critical mass transfer rate, under which the disk accretion is
  unstable. The mass transfer rate has been slightly smoothed for
  clarity. }
\label{Fig2}
\end{figure}

Firstly, we calculated the evolutionary sequences of an LMXB with
$M_{2,\rm i}=1.5M_{\odot}$ and $P_{\rm orb,i}=2.57$ d. In Figs.
\ref{Fig1} and \ref{Fig2} we plot orbital period and donor star
mass, and the mass transfer rate and the neutron star mass,
respectively. In the early phase, the orbital period decrease to
be approximately 2.56 d due to magnetic braking even if there is
no mass exchange. With nuclear evolution, the donor star fills its
Roche lobe when its age is 2.54 Gyr. When $M_{2}=0.66M_{\odot}$,
and $P_{\rm orb}=9.66$ d, the binary system detaches and mass
transfer ceases about 4.9 Myr. This detached phase also is found
in other studies for the evolution of LMXBs
(\cite{taur99,han00,pods02}). After the temporary shrinkage, the
donor star fills its Roche lobe again and climbs the giant branch.
When the age is 2.79 Gyr, the second Roche lobe overflow (RLOF)
episode terminates. Hereafter, the donor star evolves to a helium
WD and cools. The evolution end point is a binary pulsar with an
orbital period of 68.3 d. The binary consists of an He-core WD of
$0.3070M_{\odot}$ and a neutron star of $1.5184M_{\odot}$. Because
the material is transferred from the more massive donor star to
the less massive neutron star, the mass transfer proceeds
initially on short thermal timescale about 10 Myr. In this phase,
the mass transfer rate is in a range of $10^{-7}$ to
$10^{-8}M_{\odot}\rm yr^{-1}$, and the LMXB is a short-lived
persistent X-ray source. Subsequently, the mass transfer occurs on
a much longer nuclear timescale of 0.25 Gyr, with a rate of
$10^{-9}$ or $10^{-8}M_{\odot}\rm yr^{-1}$. This rate is always
less than the critical mass transfer rate. Therefore, the
progenitor of PSR J1713+0747 should be a transient X-ray source
about 95\% of all its life.

\begin{figure}
\centering \includegraphics[angle=0,width=9cm]{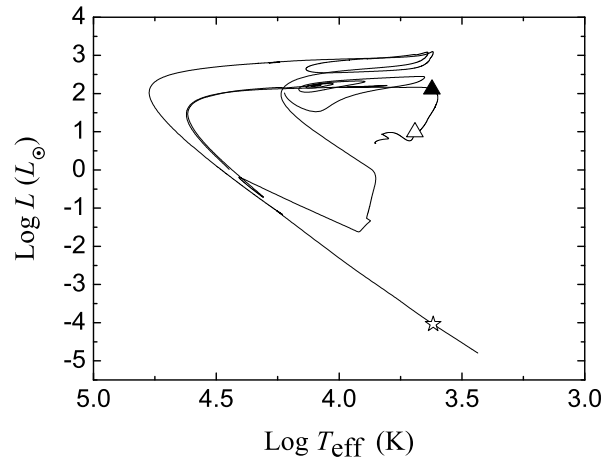}
\caption{Calculated evolutionary tracks in H - R diagram for the donor
  star with an initial mass of $1.5M_{\odot}$. The open triangle,
  solid triangle, and open star indicate the beginning of RLOF, the
  end of the last episode of RLOF and the present situation of the
  donor star, respectively. }
\label{Fig3}
\end{figure}

\begin{figure}
\centering
\includegraphics[angle=0,width=9cm]{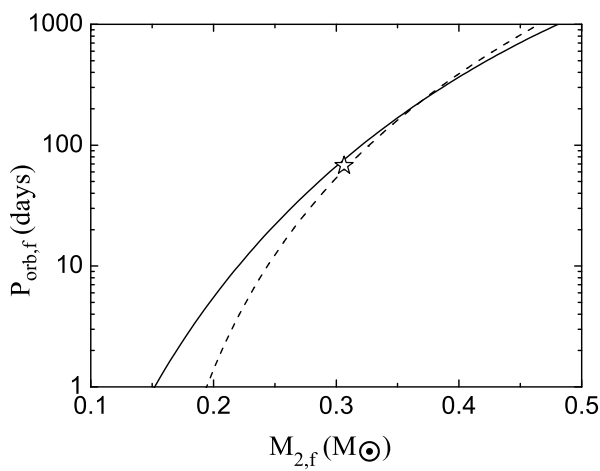}
\caption{Theoretical predicted relation between $P_{\rm orb,f}$ and
  $M_{2,\rm f}$ for binary radio pulsars. The solid curves and dashed
  curves correspond to the relations obtained by \cite{rapp95} and
  \cite{taur99}. The open star sign represents our simulation
  result. }
\label{Fig4}
\end{figure}

In Fig. \ref{Fig3}, we plot the evolutionary path of the donor
star in the Hertzsprung-Russell diagram (HRD). After the donor
star evolves into a pre-WD, a hydrogen thermonuclear flash occurs.
During this stage the element diffusion is a key physical
ingredient in obtaining a thin hydrogen envelope. We stress that
thermonuclear flashes also occur in the absence of element
diffusion, but in this case the residual hydrogen envelope is
thick, according to \cite{benv04} and \cite{pane07}. Fig.
\ref{Fig4} presents theoretical predicted relation between $P_{\rm
orb,f}$ and $M_{2,\rm f}$ for binary radio pulsars. It is clear
that our simulation result is consistent with the theoretical
prediction.

To investigate the distribution of the initial donor star mass and
orbital period for the progenitor system of PSR J1713+0747, we
calculated the evolution of a large amount of LMXBs for different
initial input parameters. Table 1 summarizes the calculated
results. Assuming that the cooling timescale of the WD is
$\tau_{\rm
  cool}=8$ Gyr, as seen in Table 1 makes it clear that the donor star
with an initial mass of $M_{2,\rm i}=1.0-1.2M_{\odot}$ would
require times in excess of the Hubble time in order to evolve into
a dim WD like the one observed in PSR J1713+0747. Therefore, PSR
J1713+0747 may originated from an LMXB with a donor star of
$M_{2,\rm
  i}=1.3-1.6M_{\odot}$ and an initial orbital period of $P_{\rm
  orb,i}=2.40 - 4.15$ d. Thus with an age of $\tau_{\rm cool}=8$
  Gyr, we obtain WD effective temperatures in the range of $\sim
3870-4120$ K, approximately in agreement with the observational
datum 3700$\pm$100 K given by \cite{lund96}, \cite{benv06} derived
a value of $4320\pm180$ K and $4250\pm250$ K determined by
\cite{scho00}. In addition, the luminosity of the WD was
determined to be $-4.171\leq {\rm Log}(L/L_{\odot})\leq -4.047$.
Apparently, the final effective temperature of the WD in our
calculations is sensitive to its cooling timescale, and a long
cooling timescale can result in a low effective temperature. On
the basis of these models with real history, we estimated the
range of surface gravity for the WD in $7.386\leq {\rm Log}(g)\leq
7.401$ ($g$ in units of g cm$^{-2}$). Detailed tabulations of our
WD evolutionary sequences are available at
http://www.fcaglp.unlp.edu.ar/$\sim$panei/papers.html.

Although the final mass of the neutron star (see Nos. 6 and 7 in
Table 1) is slightly higher than the one derived by \cite{spla05},
it is still in the allowed mass interval by theoretical orbital
period - WD mass relation ($M_{1}=1.53^{+0.08}_{-0.06}M_{\odot}$).
Though adopting various scenarios, our results are approximately
consistent with the ones derived by \cite{benv06}.

\begin{table*}
 \centering
 \begin{minipage}{140mm}
  \caption{Calculated results for different initial input
    parameters. }
\begin{tabular}{@{}lllllllllllll@{}}
  \hline\noalign{\smallskip}
 No. & $M_{\rm 2,i}$ & $P_{\rm orb,i}$ & $t_{0}$ & $t_{\rm f}$ & $t_{\rm f}-t_{0}$ & $M_{\rm 2,f}$ & $M_{1,\rm f}$ & $P_{\rm orb, f}$ & $t_{\rm p,8}$ & ${\rm Log}L_{8}$ & $T_{\rm eff,8}$ & ${\rm Log}g_{8}$\\
$$ & $(M_{\odot})$ & (days) & $(\rm Gyr)$ & $(\rm Gyr)$ & $(\rm Gyr)$ & $(M_{\odot})$ & $(M_{\odot})$ & (days) & (Gyr) & ($L_{\odot}$) & (K) & (cm s$^{-2}$)\\
 \hline\noalign{\smallskip}
1 & 1.0 & 5.65 & 12.48 & 12.68 & 0.20 & 0.3058 & 1.4211 & 66.7 & 20.68 & -4.0318 & 4152 & 7.3810 \\
2 & 1.1 & 5.09 &  8.74 &  8.95 & 0.21 & 0.3069 & 1.4220 & 68.8 & 16.95 & -4.1054 & 3999 & 7.3909 \\
3 & 1.2 & 4.48 &  6.12 &  6.34 & 0.22 & 0.3063 & 1.4247 & 67.6 & 14.34 & -4.1349 & 3937 & 7.3925 \\
4 & 1.3 & 4.15 &  4.70 &  4.92 & 0.22 & 0.3067 & 1.4260 & 68.1 & 12.92 & -4.1114 & 3985 & 7.3905 \\
5 & 1.4 & 3.70 &  3.30 &  3.54 & 0.24 & 0.3068 & 1.4964 & 68.0 & 11.54 & -4.1707 & 3872 & 7.4001 \\
6 & 1.5 & 2.57 &  2.54 &  2.79 & 0.25 & 0.3070 & 1.5184 & 68.3 & 10.79 & -4.0471 & 4122 & 7.3856 \\
7 & 1.6 & 2.40 &  1.96 &  2.20 & 0.24 & 0.3068 & 1.5383 & 67.5 & 10.20 & -4.1682 & 3880 & 7.4010 \\
 \noalign{\smallskip}\hline
\end{tabular}
\note{The meaning of the columns are presented as follows.
    $M_{\rm 2,i}$: initial mass of the donor star; $P_{\rm orb,i}$:
    initial orbital period; $t_{0}$: age when Roche lobe overflow
    begins; $t_{\rm f}$: age when the main Roche lobe overflow ends;
    $t_{\rm f}-t_{0}$: mass transfer timescale; $M_{\rm 2,f}$: mass of
    the WD; $M_{\rm 1,f}$: mass of the neutron star; $P_{\rm
      orb,f}$: final orbital period; $t_{\rm p,8}$, ${\rm Log}L_{8}$,
    $T_{\rm eff,8}$ and ${\rm Log}g_{8}$ : age, luminosity, effective
    temperature and gravity when $\tau_{\rm co}=8$ Gyr,
    respectively.}
\end{minipage}
\end{table*}

\section[]{Summary and discussion}

In this paper, we attempt to carefully study the evolution history
and the progenitor characteristics of PSR J1713+0747. Based on
Eggleton's stellar evolution code, we calculated the evolutionary
sequences of a number of LMXBs to fit the observed data ($M_{1},
M_{2}$, and $P_{\rm orb}$) of PSR J1713+0747. For the WD regime we
employed the stellar evolution code {\tt LPCODE} (\cite{alth05}).

In calculations, angular momentum loss via magnetic braking is
included if the mass of the donor star is less than
$1.5M_{\odot}$. The observations show that PSR J1713+0747 has only
accreted less than $0.1M_{\odot}$ from the donor star during
evolution. Unstable mass transfer due to the thermal and viscous
instability of the accretion disk may be responsible for the
extensive mass loss \footnote{Certainly, the propeller effect of
the pulsar cannot be excluded (\cite{illa75})}. Therefore, LMXBs
may be transient X-ray sources for most of their life. If we adopt
a duty cycle of 0.01, the progenitor of PSR J1713+0747 can be
constrained to be an LMXB with an initial donor star of $1.3-1.6
M_{\odot}$ (Z=0.02) and an initial orbital period of $2.40 -4.15$
d. Adopting a cooling timescale of 8 Gyr for the WD, our
calculations show that at present its effective temperature is
between 3870 and 4120 K. Compared with the results derived by
\cite{benv06}, for Z=0.02, the cooling timescale of the WD in our
model is close to the spin-down age of the
pulsar\footnote{\cite{benv06} propose that the WD needs $10.8\pm
2$ Gyr to cool when Z=0.02.}. We obtained absolute magnitudes of
$M_V \approx$15.7$\pm$0.4 and $M_I \approx$13.8$\pm$0.3 in the $V$
and $I$ bands, and inferred effective temperatures $T_{\rm eff}
\approx$ 3301 - 4169 K and $T_{\rm eff} \approx$ 3946 - 4694 K.
Detailed tabulations of our WD evolutionary sequences are
available at
http://www.fcaglp.unlp.edu.ar/$\sim$panei/papers.html.

Our accretion disk instability model successfully reproduces the
orbital period and the mass of the WD. However, the effective
temperature of the WD is slightly higher than observations,
although we consider gravitational settling and chemical and
thermal diffusion during the evolution of the WD (Panei et al.
2007). Employing \emph{Hubble Space Telescope} (\emph{HST}),
\cite{lund96} performed optical observations for the WD companion
of PSR J1713+0747. Adopting the temperature calibration of ($V -
I$ ) colour given by \cite{mone92}, they presented the effective
temperature of the WD is $T_{\rm eff}=3700\pm 100$ K. Assuming a
blackbody spectrum, \cite{hans98} fitted the \emph{HST} data, and
obtained $T_{\rm eff}=3430\pm 270$ K. A long cooling timescale of
the WD may be responsible for the discrepancy between our
evolution results and the inferred value of observations. In this
work we take $\tau_{\rm co}=8$ Gyr (\cite{spla05}).  However, it
is very rough to estimate the spin-down age of the pulsar from
$\tau_{\rm PSR}=P/2\dot{P}$ , where $P$ and $\dot{P}$ are the spin
period and the period derivative of the pulsar, respectively
\footnote{The standard spin-down age of the pulsar is
  $\tau_{\rm PSR}=\frac{P}{(n-1)\dot{P}}[1-(P_{0}/P)^{n-1}]$, where
  $P_{0}$ is the initial period and $n$ the braking index of the
  pulsar. When $P_{0}\ll P$ and $n=3$, $\tau_{\rm
    PSR}=P/2\dot{P}$. However, the derived braking indice of several
  young radio pulsars are all less than 3 (\cite{xu01,chen06b}).}.  We
expect that further spectroscopic and photometric observations can
help us understand the evolutionary track of PSR J1713+0747.

As an alternative evolution channel, millisecond pulsars may also
be formed by accretion-induced collapse (AIC) when the mass of the
accreting ONe WD exceeds the Chandrasekhar mass of $1.4M_{\odot}$
(\cite{nomo84,nomo86}).  Statistical analysis shows that the
birthrate of LMXBs is 1-2 orders of magnitude lower than that of
millisecond pulsars (\cite{cote89,lori95}). The AIC channel was
raised to interpret this birthrate discrepancy (\cite{bail90}).
Based on a scenario including both the AIC of an accreting WD and
unstable disk accretion, \cite{xu09} propose that GRO J1744-28 may
be evolved from a WD binary. Recently, the estimated birthrates of
millisecond pulsars show that the often neglected AIC channel
cannot be ignored (\cite{hurl10}). Therefore, the formation
possibility of PSR J1713+0747 via the AIC of a massive WD cannot
be ruled out. We plan to explore the AIC channel for the formation
and evolution of PSR J1713+0747.

Of course, there are many uncertainties in our input physics.
Firstly, the orbital angular momentum loss plays a vital role in
the evolution of LMXBs. Some mechanism of extracting angular
momentum possibly exists besides magnetic braking and mass loss
due to the accretion disk instability. For example, the propeller
effect of the neutron star may carry away the specific orbital
angular momentum at the magnetosphere of the neutron star
(\cite{illa75}). In addition, a circumbinary disk around LMXBs may
also extract orbital angular momentum efficiently from the binary
system (\cite{spru01,taam01}). Secondly, our results strongly
depend on duty cycles, which are poorly known. In this work, we
take a constant duty cycle. However, duty cycles may change from
one system to another (\cite{laso01}), and change with the orbital
period and the mass transfer rate. Finally, a gamma-ray flux
caused by the interaction of the magnetic field of the neutron
star with the accretion disk may drive a wind from the donor star
by heating its outer layers; i. e., the donor star may be
evaporated (\cite{heuv88,kluz88,rude89a,rude89b}).

\begin{acknowledgements}
We are grateful to the referee, Christopher A. Tout, for his
constructive suggestions for improving this manuscript. We also
thank Zhan-Wen Han and Xue-Fei Chen for their helpful discussions,
and thank Stephen Justham for his help in improving the English of
this manuscript. This work has been supported in part by the
National Natural Science Foundation of China (Grant No 10873011),
Program for Science \& Technology Innovation Talents in
Universities of Henan Province, and Innovation Scientists and
Technicians Troop Construction Projects of Henan Province, China.
We acknowledge the Facultad de Ciencias Astron\'omicas y
Geof\'{\i}sicas (UNLP), Instituto de Astrof\'{\i}sica de La Plata
(CCT La Plata, CONICET-UNLP) and Lic. H. Viturro for his technical
support.
\end{acknowledgements}

\end{document}